\documentclass[prb,aps,twocolumn,showpacs]{revtex4-1}
\usepackage{amsfonts,amsmath,bm,multirow}
\usepackage[OT4]{fontenc}

\usepackage{graphicx}
\usepackage{xcolor}
\usepackage{epstopdf}
\usepackage{dsfont}

\newcommand{\rr}{{\bm{r}}}
\newcommand{\kp}{\bm{k}\!\cdot\!\bm{p}}

\newcommand{\cp}{{\mathrm{c.p.}}}

\newcommand{\rl}{\rangle\!\langle}

\newcommand{\ket}[1]{\left| #1 \right\rangle} 

\newcommand{\Tr}[1]{\mathrm{Tr}\!\left[#1\right]}

\begin{document}

\title{Spin-orbit coupling and magnetic field dependence of carriers states in \\a self-assembled quantum dot}
\author{Krzysztof Gawarecki}
\email{Krzysztof.Gawarecki@pwr.edu.pl} 
\affiliation{Department of Theoretical Physics, Faculty of Fundamental Problems of Technology, Wroc\l aw University of Technology, Wybrze\.ze Wyspia\'nskiego 27, 50-370 Wroc\l aw, Poland}

\begin{abstract}
	In this work I investigate the influence of spin-orbit coupling on the magnetic field dependence of carrier states in a self-assembled quantum dot. I calculate the hole energy levels using the $6$, $8$ and $14$ band $\kp$ model. Through a detailed study within these models, I extract the information about the impact of various spin-orbital coupling channels in the hole $p$-shell.  I also show that complicated magnetic field dependence of the hole $p$-shell resulting from numerical simulations, can be very well fitted using a phenomenological model.  I compare the electron and hole g-factors calculated within $8$ and $14$ band $\kp$ models and show that these methods give reasonably good agreement. 
\end{abstract}


\maketitle

\section{Introduction}
\label{sec:intr}

Spin-related properties of carriers confined in quantum dots (QDs) attract much attention due to their potential application in spintronics\cite{loss98,recher00,yong13}.
In particular, the effect of spin-orbit (SO) coupling in QDs was a subject of extensive studies over recent years\cite{rodriguez04,bulaev05b,Manaselyan09,Vachon09,Siranush12,Siranush13}. The lack of inversion symmetry in the crystal lattice (bulk inversion asymmetry, BIA) leads to the Dresselhaus SO coupling, while asymmetry in the shape of a nanostructure (or induced via external fields) gives rise to Rashba SO coupling (structure inversion asymmetry, SIA)\cite{winkler03}. Furthermore, in the case of nanostructures, other mechanisms related to abrupt material interfaces play a very important role\cite{durnev14} . 
In the description of QDs spectra, some theoretical works\cite{rodriguez04,bulaev05b,Siranush12,Siranush13} utilize the effective Fock-Darwin model supplemented by additional terms (described by empirical parameters) which represents Rashba or/and Dresselhaus SO coupling. Such a model qualitatively describes basic QD properties and  (if taken with realistic parameters) could predict correct spin ordering of the several lowest carrier states. The spin dynamics of a QD system can be driven by applying an external magnetic field. However, the system response strongly depends on the parameters which cannot be simply deduced from the bulk values. In consequence, spin control needs a precise knowledge of the carrier states in the system, which requires advanced modeling. One of the important issues is related to the electron/hole $g$-factor, which in the case of QD can differ by the one order of magnitude to the bulk values\cite{pryor06}. The $g$-factors of electron/hole in a QD were investigated in many experimental works\cite{medeiros02,Vachon09,kleemans09,klotz10,schwan11,nakaoka04b,andlauer09,pryor12,jovanov12} and theoretical studies including tight-binding \cite{sheng07,sheng08} and $\kp$ \cite{nakaoka04b,andlauer09,pryor12,jovanov12} modeling. 

In this paper I study the magnetic field dependence of the electron/hole states. I utilize the $14$ band $\kp$ model, which inherently accounts for Rashba, Dresselhaus and other spin-orbit coupling mechanisms. 
I demonstrate that a pronounced anticrossing emerging in the hole spectrum, results from the interplay of several SO mechanisms with the most important contribution from the one that depends on the shear strain. I also show that the results from the $14$-band $\kp$ theory can be very well fitted via an extended Fock-Darwin model with the structural asymmetry and SO interaction taken into account. Finally, I calculate the electron and hole g-factors using $8$ and $14$ band $\kp$ models and show that these methods give reasonably good agreement.

The paper is organized as follows. In Sec.~\ref{sec:model}, I briefly describe the models which are used to calculate strain distribution and the carrier states. In Sec.~\ref{sec:results}, I discuss the results of numerical simulations. Sec.~\ref{sec:conclusion} contains concluding remarks. Finally, a detailed description of the model is given in the Appendix.

\section{Model}
\label{sec:model}

\begin{figure}[tb]
	\begin{center}
		\includegraphics[width=80mm]{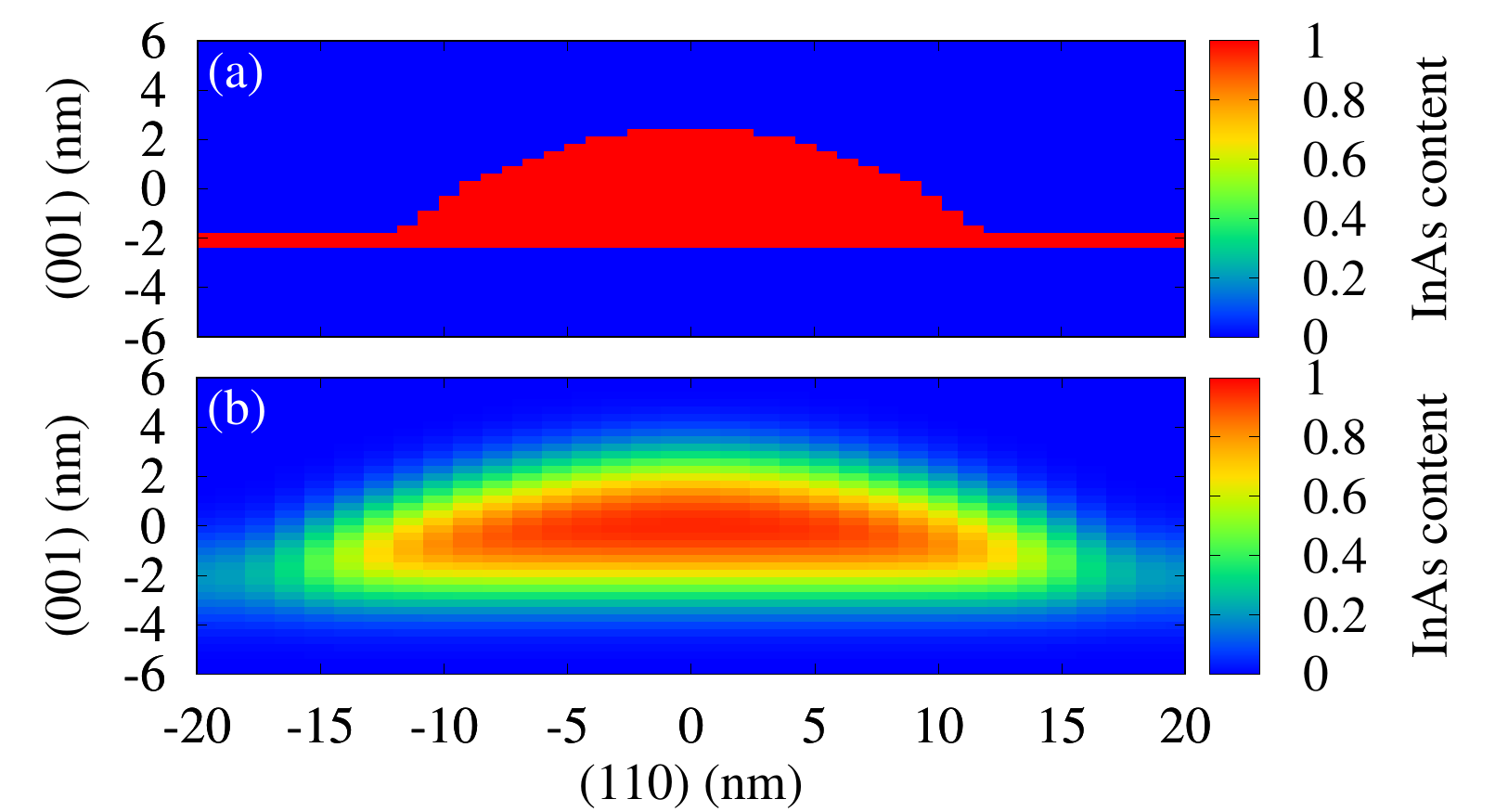}
	\end{center}
	\caption{\label{fig:comp}\textcolor{gray}({Color online) Material distribution in the system, in the case of uniform (a) and blurred (b) QD. } }
\end{figure}
The system under consideration contains a single self-assembled QD formed by $\mathrm{In} \mathrm{Ga} \mathrm{As}$ in GaAs matrix. The dot is placed on a $0.6$~nm thick wetting layer (WL). The dot is lens-shaped  with the diameter $24$~nm and the height $4.2$~nm. I consider the uniform (Fig.~\ref{fig:comp}(a)) and blurred (Fig.~\ref{fig:comp}(b)) $\mathrm{In}_{x}\mathrm{Ga}_{1-x}\mathrm{As}$ compositions. In the latter case, the material distribution is processed by gaussian blur with standard deviation of $1.2$~nm. The inhomogeneous composition is described by the function $C(\rr)$, where $C(\rr)=1$ refers to pure InAs, $C(\rr)=0$ to pure GaAs and values between correspond to $\mathrm{In}_{x}\mathrm{Ga}_{1-x}\mathrm{As}$.
The spectral properties of the system are affected by strain which is caused by InAs/GaAs lattice mismatch. The strain distribution is modeled within continuous elasticity approach \cite{pryor98b}. 
The piezoelectric potential $V(\rr)$ is calculated by solving the equation $ \rho(\rr)  = \varepsilon_{0} \nabla (  \varepsilon_{r}(\rr) \nabla V(\rr) )$,
where  $\varepsilon_{r}(\rr)$ is the position-dependent relative permittivity. The charge density is calculated from $\rho(\rr) = -\nabla \cdot \bm{P}(\rr)$,
where the piezoelectric polarization is accounted for up to the second order with respect to the strain tensor elements\cite{bester06a}. 

In order to calculate the electron and hole states, I implemented the $14$ band $\kp$ model. In this framework (in contrast to $8$ and fewer band approaches) the kinetic part of the Hamiltonian correctly describes the symmetry ($C_{\mathrm{2v}}$) of the zinc-blende crystal\cite{tomic11}. Furthermore, the model inherently contains Dresselhaus, Rashba and other coupling mechanisms which, in fewer-band models, need to be represented via additional perturbative terms\cite{winkler03}. The Hamiltonian of the system can be written as $H = H^{\mathrm{k}} + H^{\mathrm{str}} + H^{\mathrm{m}} + V $, where $H^{\mathrm{k}}$ is the kinetic part of the Hamiltonian, $H^{\mathrm{str}}$ accounts for the strain and $H^{\mathrm{m}}$ represents the magnetic interaction. The hamiltonian can be divided into blocks according to the symmetry classification\cite{mayer91,winkler03}. In the case of the extended Kane model ($14$ bands),  $\Gamma_{8c}$, $\Gamma_{7c}$ $\Gamma_{6c}$, $\Gamma_{8v}$, and $\Gamma_{7v}$ bands are taken into account explicitly and the remote band contribution enters via material parameters\cite{mayer91,winkler03}. I keep the Burt-Foreman operator ordering \cite{foreman97} in its extended version proposed in Ref.~\onlinecite{eissfeller12}.  The magnetic field is introduced via Peierls substitution within gauge invariant scheme, described in detail in Ref.~\onlinecite{andlauer08}. The electron and hole states are obtained by diagonalizing the Hamiltonian. The in-plane probability density of the $n$-th electron/hole state is calculated using the formula
\begin{equation*}
	d^{(\mathrm{e/h})}_{n}(x,y) = \sum_{m=1}^{14} \int_{-\infty}^{\infty} \psi^{(\mathrm{e/h})*}_{n,m}(x,y,z) \psi^{(\mathrm{e/h})}_{n,m}(x,y,z) dz,
\end{equation*} 
where $\psi^{(\mathrm{e/h})}_{n,m}(x,y,z)$ denotes $m$-th band component of the $n$-th electron/hole wavefunction. The average values of the $z$-components of the electron and hole envelope angular momenta read
\begin{align*}
	\langle M_{z} \rangle &= \sum_{m=1}^{14} \int_{-\infty}^{\infty} \psi^{(\mathrm{e/h})*}_{n,m}(\rr) \left( i y \partial_{x} - i x \partial_{y}  \right)  \psi^{(\mathrm{e/h})}_{n,m}(\rr) d \rr.
\end{align*} 
A full description of the Hamiltonian and all of the calculation details are given in the Appendix. 

\section{Results}
\label{sec:results}

This Section presents the results for the magnetic field dependence of the lowest electron and hole states. The eigenstates are $14$-element spinors, where each part refers to the one of the bands $\Gamma_{8c}$, $\Gamma_{7c}$ $\Gamma_{6c}$, $\Gamma_{8v}$, or $\Gamma_{7v}$. Since the considered QD has geometrical axial symmetry, the states are denoted according to the axial components of their approximate \textit{envelope} angular momenta ($s$, $p$, $d$ shells). 

\subsection{Electron states}
\begin{figure}[tb]
	\begin{center}
		\includegraphics[width=3.5in]{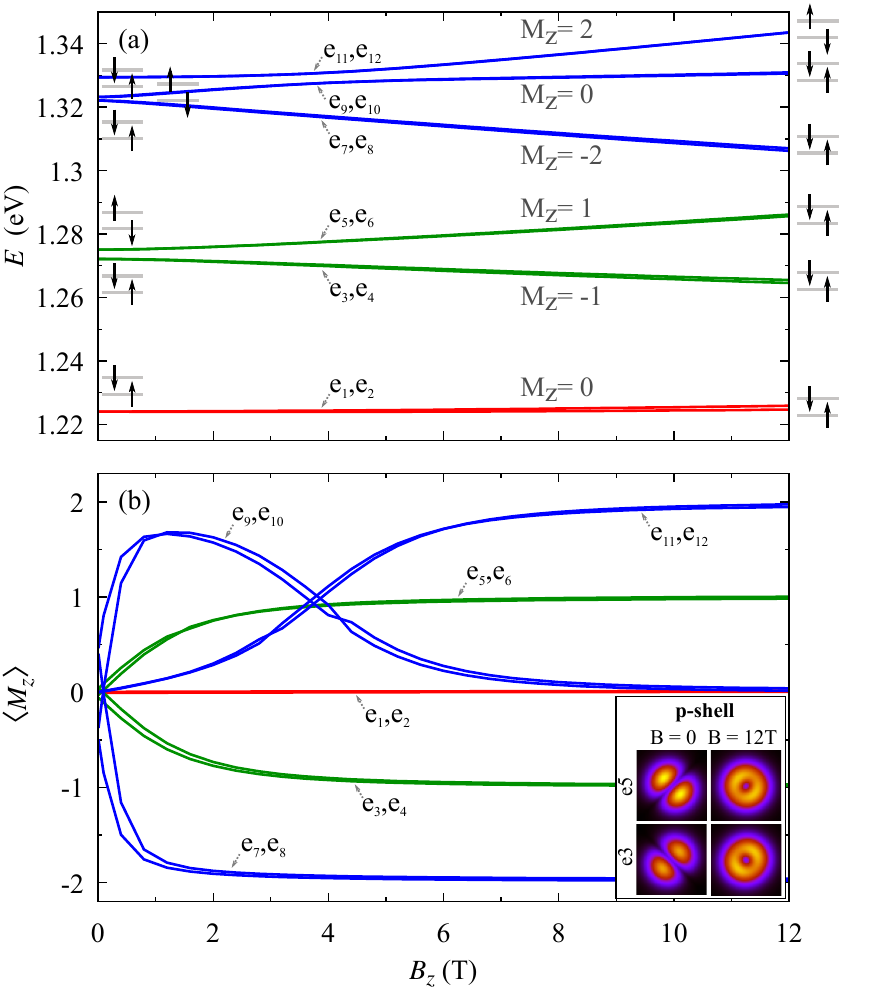}
	\end{center}
	\caption{\label{fig:mag_el}\textcolor{gray}({Color online) (a) Magnetic field dependence of the $12$ lowest electron states in the uniform InAs QD. Energy $E=0$ refers to the unstrained GaAs valence band edge. Schematic pictures on the left describes spin configuration at magnetic field close to zero, while insets on the right presents the configuration at B=$12$~T. (b) Corresponding  axial projection of the envelope angular momentum. The inset in the right bottom corner presents the in-plane probability density of $e_{3}$ and $e_{5}$ states at $B_{z}=0$ and $B_{z}=12$~T. } }
\end{figure}
The electron energy levels in the uniform InAs QD as a function of external axial magnetic field are shown in Fig.~\ref{fig:mag_el}a. The corresponding average values of the axial projection of the envelope angular momentum $\langle M_{z} \rangle $ are presented in Fig.~\ref{fig:mag_el}b. The two lowest states ($e_{1},e_{2}$) are $s$-type with $\langle M_{z} \rangle \approx 0$. Their energies increases according to the diamagnetic shift $\alpha^{(\mathrm{e})}_{s} B_{z}^{2}$, where fitting to the numerical data results in $\alpha^{(\mathrm{e})}_{s} = 8.68$~$\mathrm{\mu eV/T^{2}}$. The  energy splitting between the two lowest $s$-states can be attributed to the effective Land\'e factor $g_{\mathrm{e}} = (E_{s \uparrow} - E_{s \downarrow})/ \mu_{B} B_{z} = -1.77$. This value differs significantly from the  electron  bulk g-factor, averaged according to the local composition $$ \langle g \rangle = \sum_{m=1}^{14} \int_{-\infty}^{\infty} \psi^{(\mathrm{e})*}_{s,m}(\rr) g(\rr) \psi^{(\mathrm{e})}_{s,m}(\rr) d \rr = -13.1,$$ where $g(\rr) = C(\rr) g_{\mathrm{InAs}} + (1-C(\rr)) g_{\mathrm{GaAs}} $,  with $g_{\mathrm{InAs}} = -14.9$ and $g_{\mathrm{GaAs}} = -0.44$. This discrepancy is caused by the renormalization of the effective gap and the angular momentum quenching\cite{pryor06}. In fact, it has been shown\cite{pryor06} that the Land\'e factor $g_{\mathrm{e}}$ tends to the bulk value in the limit of a very large dot and to $g_{0}=2$ in the limit of a very small dot with strong confinement. The next four states ($e_{3}$ -- $e_{6}$, green lines in Fig.~\ref{fig:mag_el}a) exhibit the $p$-type symmetry. The splitting at $B=0$ is caused by the piezoelectric field and (with smaller contribution) by the spin-orbit coupling. The pairs of these states exhibit large Zeeman splitting due to their nonzero envelope angular momenta. As shown in Fig.~\ref{fig:mag_el}b (inset), in weak magnetic field the $p$ states are mostly oriented along ($110$) and ($1\bar{1}0$) axes. In consequence, their $\langle M_{z} \rangle $ is close to zero. However, the magnetic field leads to the mixing of these states and tends to formation of $\ket{p_{\pm 1}} = (\ket{p_{110}} \pm i \ket{p_{1\bar{1}0}})/\sqrt{2} $ with $M_{z}=\pm 1$, which are clearly visible at $B_{z} > 4$~T. 
Finally, the last three pairs of states belongs to the $d$ shell ($e_{7}$ -- $e_{12}$, plotted with blue lines in Fig.~\ref{fig:mag_el}a), for sufficiently strong magnetic field they form $M_{z} \approx -2,0,2$ configurations.  At about $B=4$~T there is an anticrossing between the state with $M_{z} \approx 0$ and $M_{z} \approx 2$. All of the considered electron states are mainly $\Gamma_{6c}$ with a very small admixture from the other groups of bands. In consequence their axial projection of the \textit{band} angular momentum  $j^{(\mathrm{e})}_{z}$ is approximately $\pm 1/2$.  

Although the structure of the electron energy levels resembles the standard Fock-Darwin model, the obtained spin configuration is more complicated. The reason is the spin-orbit coupling, which affects the spin and spectral properties of a QD \cite{rodriguez04}. The total spin-orbit coupling favors the configurations where the electron spin and envelope angular momentum are antiparallel. In consequence, for positive $M_{z}$ and weak magnetic fields, the spin configuration is inverted compared to the $s$-shell states (left spin diagrams in Fig.~\ref{fig:mag_el}a). However at stronger magnetic fields, the energy related to the Zeeman term dominates and the spin configuration returns to the ,,usual" case where the lower state is oriented up and the higher down (spin diagrams on the right in Fig.~\ref{fig:mag_el}a). This situation takes place for $e_{5}$, $e_{6}$ pair of states ($M_{z}\approx 1$): at about $B_{z}=1.2$~T there is a crossing between these states. However, in the case of $M_{z} \approx 2$ states, the considered range of the magnetic field is too weak for this effect to occur.

\begin{figure}[tb]
	\begin{center}
		\includegraphics[width=3.5in]{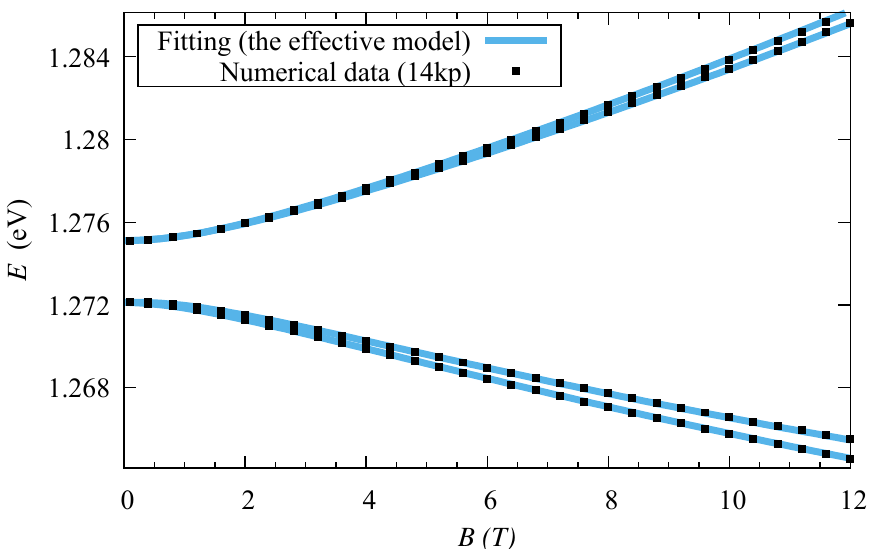}
	\end{center}
	\caption{\label{fig:mag_p_el}\textcolor{gray}({Color online) Magnetic field dependence of the electron $p$-shell energy levels in the uniform InAs QD. } }
\end{figure}
\begin{table}
	\begin{tabular}{|c|c|c|}
		\hline 	
		& Uniform QD & Blurred QD \\
		\hline 	
		$ V^{(\mathrm{e})}_{\mathrm{a}}$ $[$meV$]$						 					& $1.49$  & $0.597$ \\
		\hline 
		$V^{(\mathrm{e})}_{\mathrm{so}}$ $[$meV$]$      				                    & $0.133$ & $0.296$ \\
		\hline
		$g^{(\mathrm{e})}_{\mathrm{p}}$            						 					& $-1.088$ & $-0.558$ \\
		\hline 
		$W^{(\mathrm{e})} \left [\mathrm{ {meV }/{T}} \right ]$ 					& $0.861$ &  $0.840$\\
		\hline 
		$\alpha^{(\mathrm{e})}_{\mathrm{p}}\left [\mathrm{ {\mu eV }/{T^2}} \right ]$  	& $12.85$ &	$11.93$ \\
		\hline 
	\end{tabular} 
	\caption{\label{tab:param_fit_el}The effective parameters describing the electron $p$-shell obtained from the fitting procedure.}
\end{table}
The complicated magnetic field dependence of the electron $p$-shell can be interpreted within the phenomenological model involving $5$ parameters. This corresponds to the Fock-Darwin model with spin-orbit term included \cite{rodriguez04}. In the basis of four states  $ \left \{ \ket{+1 \uparrow},\ket{-1 \uparrow},\ket{+1 \downarrow},\ket{-1 \downarrow} \right \}$ (where the first index refers to $M_{z}$ and the second one to the spin orientation), the relevant Hamiltonian is
\begin{align}
  \label{el_effm}
	H^{(\mathrm{e})}_{\mathrm{p}} &= V^{(\mathrm{e})}_{\mathrm{a}} \left( | +1 \rl -1|  + \mathrm{h.c.}  \right) \otimes \mathds{1} + \frac{1}{2 \hbar}  V^{(\mathrm{e})}_{\mathrm{so}} L_{z} \otimes \sigma_{z} \nonumber \\ 
	&\phantom{=}  +  \frac{1}{2}  \mu_{B} g^{(\mathrm{e})}_{\mathrm{p}} B_{z} \mathds{1} \otimes  \sigma_{z} + \frac{1}{\hbar} W^{(\mathrm{e})} B_{z} L_{z} \otimes \mathds{1}   \nonumber  \\ &\phantom{=}+ \alpha^{(\mathrm{e})}_{\mathrm{p}} B^{2}_{z} \mathds{1} \otimes \mathds{1},
\end{align}
where the tensor product is related to the formal subsystems of the envelope and the band angular momentum,  $V^{(\mathrm{e})}_{\mathrm{a}}$ is a parameter which accounts for the system anisotropy, $\mathds{1}$ is the unit operator, $\sigma_{z}$ is the axial Pauli matrix, $L_{z}$ is the operator of the axial projection of the envelope angular momentum, $g^{(\mathrm{e})}_{\mathrm{p}}$ is the effective electron g-factor for the $p$-shell and $W^{(\mathrm{e})}$ is a parameter related to the influence of the envelope angular momentum. The resulting Hamiltonian can be written in a form of $4 \times 4$ matrix which can be diagonalized analytically (see the Appendix). I fitted the phenomenological parameters (Tab.~\ref{tab:param_fit_el}) and obtained an excellent agreement to the 14 $\kp$ numerical data (see Fig.~\ref{fig:mag_p_el}). The blurred QD gives larger spin-obit term, but smaller anisotropy compared to the uniform QD. 
Since $V^{(\mathrm{e})}_{\mathrm{so}}$ is relatively small for both of the dots, it is possible to obtain a reasonable good fit also for the standard Fock-Darwin model (which corresponds to $V^{(\mathrm{e})}_{\mathrm{so}}=0$). However, at low magnetic fields, this produces wrong spin ordering of some states   (in the case of the $p$-shell, wrong spin ordering appears at $B_{z} < 1.2$~T for the uniform QD and $B_{z} < 10$~T for the blurred QD). The overall spin-orbit coupling in the $p$-shell ($V^{(\mathrm{e})}_{\mathrm{so}}$) is caused by the interplay of the abrupt material interfaces, the Rashba coupling, shear strain in $H_{\mathrm{6c8v}}$ and $H_{\mathrm{6c7v}}$, the Dresselhaus coupling, and subband mixing.

\subsection{Hole states}
\begin{figure}[tb]
	\begin{center}
		\includegraphics[width=80mm]{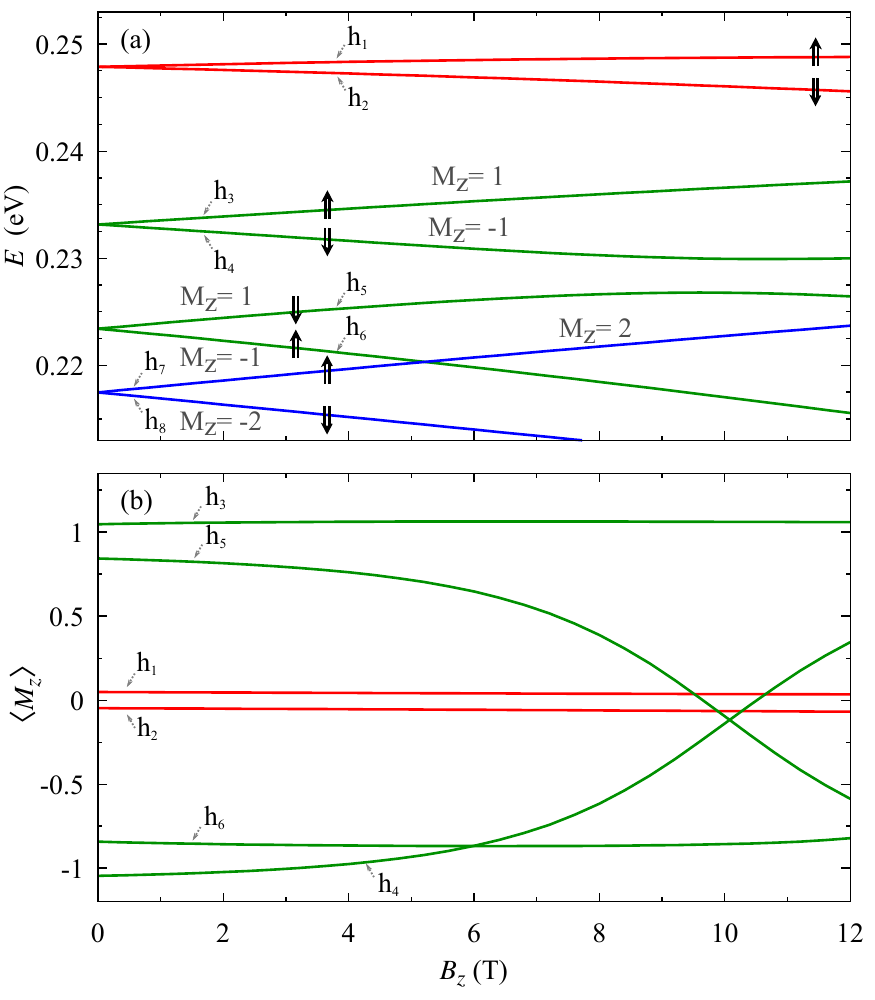}
	\end{center}
	\caption{\label{fig:mag_h}\textcolor{gray}({Color online) (a) Magnetic field dependence of the lowest hole energy levels. Energy $E=0$ refers to the unstrained GaAs valence band edge. (b) Corresponding axial projection of the envelope angular momentum.} }
\end{figure}
\begin{figure}[tb]
	\begin{center}
		\includegraphics[width=3.5in]{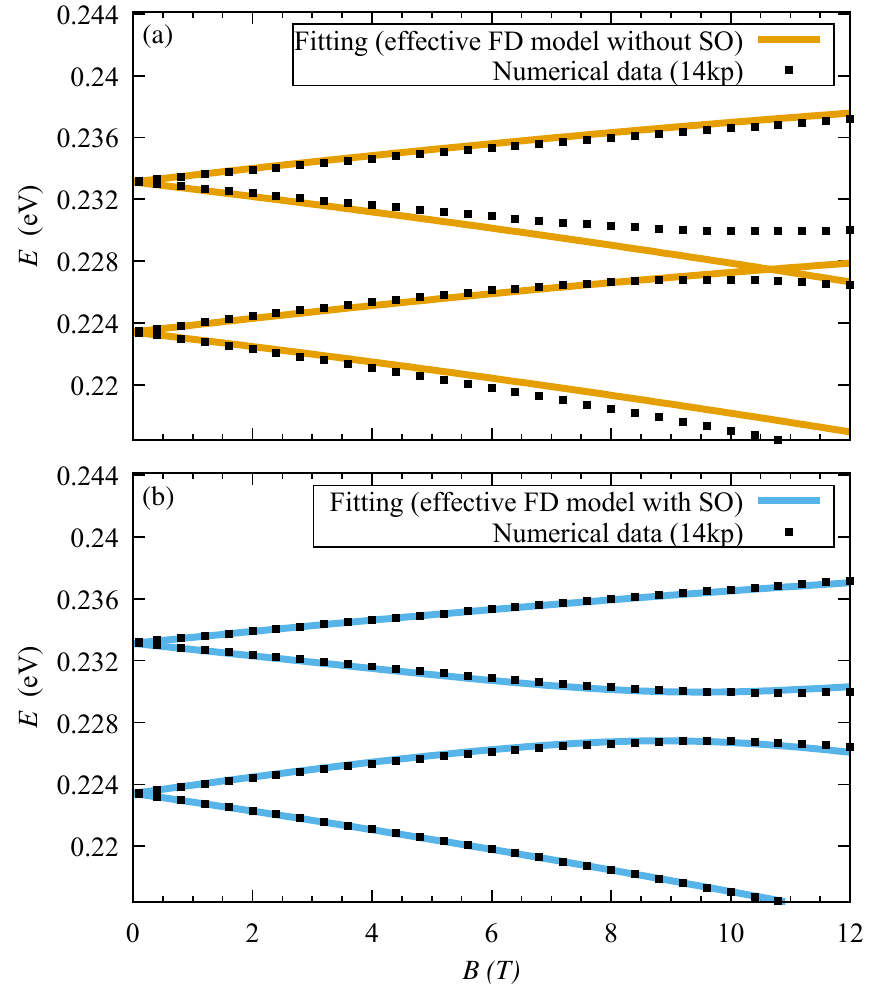}
	\end{center}
	\caption{\label{fig:mag_p_h}\textcolor{gray}({Color online) Magnetic field dependence of the hole $p$-shell energy levels in the uniform InAs QD. Solid lines denote the results obtained from the effective model for a fitting (a) without SO parameter (b) with SO parameter included. } }
\end{figure}

The valence band states (directly resulting from the numerical simulations) are converted to the hole states by the operation of time reversal, which inverts the direction of the envelope and band angular momenta.  
The magnetic field dependence of the lowest hole energy levels ($s$, $p$ and the two lowest states from the $d$ shell) is shown in Fig.~\ref{fig:mag_h}a and the corresponding axial projections of the envelope angular momenta are given in Fig.~\ref{fig:mag_h}b. Due to a complicated pattern of the higher states, the angular momenta are shown up to the $p$-shell. Contrary to the electron case, the alignment of hole energy levels does not exhibit clear shell structure\cite{williamson00,he06,schliwa07,zielinski12}.  The energy splitting between the two lowest states ($s$-type, $M_{z} \approx 0$) is significantly larger than in the electron case. The relevant effective g-factor is $g_{\mathrm{h}} =  - (E_{s \Uparrow} - E_{s \Downarrow})/ \mu_{B} B_{z} = -4.68$. Note that with this sign convention, the exciton $g$-factor is given by $g_{X} = (E_{X}(\sigma_{+}) - E_{X}(\sigma_{-}))/ \mu_{B} B_{z} = -g_{e} + g_{h}$, where the energies $E_{X}(\sigma_{\pm})$ are related to absorption of the circularly polarized ($\sigma_{\pm}$) light\cite{eissfeller12}. The diamagnetic parameter is fitted to $\alpha^{\mathrm{(h)}}_{s} = -4.67$~$\mathrm{\mu eV/T^{2}}$. 

The next four states belong to the $p$-shell. Large splitting between their spin doublets at $B=0$ arises from a combination of the spin-orbit coupling and the influence of the piezoelectric potential. Unlike the electron states, even at small magnetic field, the $M_{z} \approx \pm 1$ configurations are well defined. At about $B=10$~T, there is an anticrossing which involves the same spin orientation but different envelope angular momenta. Finally, at $B=5$~T there is a crossing between $M_z \approx -1$ state and the one from the $d$-shell (with $M_z \approx 2$). The considered $p$ states are mostly heavy holes with relatively small (below $10$\%) admixture from the other subbands, so their total axial projections of the band angular momenta can be approximated by $j^{(\mathrm{h})}_{z} \approx \pm 3/2$.

\begin{table}[tb]
	\begin{tabular}{|c|c|c|c|}
	\hline 	
	& Uniform QD* & Uniform QD & Blurred QD\\
	\hline 	
	$ V^{(\mathrm{h})}_{\mathrm{a}}$ $[$meV$]$						 					& $4.853$ & $1.581$  & $0.4787$ \\
	\hline 
	$V^{(\mathrm{h})}_{\mathrm{so}}$ $[$meV$]$      				                    & $0.0$ & $9.216$ & $7.047$ \\
	\hline
	$g^{(\mathrm{h})}_{\mathrm{p}}$            						 					& $-15.726$ & $-2.6892$ & $-2.1849$ \\
	\hline 
	$W^{(\mathrm{h})} \left [\mathrm{ {meV }/{T}} \right ]$ 					& $0.0$ & $0.4968$ &  $0.4122$\\
	\hline 
	$\alpha^{(\mathrm{h})}_{\mathrm{p}}\left [\mathrm{ {\mu eV }/{T^2}} \right ]$  	& $-6.956$ & $-6.956$ &	$-3.721$ \\
	\hline 
\end{tabular} 
	\caption{\label{tab:param_fit_h}The effective parameters for hole $p$-shell obtained from a fitting procedure, (*) denotes a fitting without the SO parameter.}
\end{table}

Similarly to the electron case, the magnetic field dependence of the $p$-states can be described using an effective model based on the empirical parameters. Then, analogously to Eq.~(\ref{el_effm}) the Hamiltonian can be written as
\begin{align*}
H^{\mathrm{(h)}}_{\mathrm{p}} &= V^{\mathrm{(h)}}_{\mathrm{a}} \left( | +1 \rl -1|  + \mathrm{h.c.}  \right) \otimes \mathds{1} + \frac{1}{2 \hbar}  V^{\mathrm{(h)}}_{\mathrm{so}} L_{z} \otimes \sigma_{z} \nonumber \\ 
&\phantom{=}  +  \frac{1}{2}  \mu_{B} g^{\mathrm{(h)}}_{\mathrm{p}} B_{z} \mathds{1} \otimes  \sigma_{z} + \frac{ 1}{\hbar} W^{\mathrm{(h)}} B_{z} L_{z} \otimes \mathds{1} + \alpha^{\mathrm{(h)}}_{\mathrm{p}} B^{2}_{z}.
\end{align*}
Fig.~\ref{fig:mag_p_h} presents $14$~$\kp$ results of the $p$-shell energy branches compared to the effective model in two cases: (a) the fitting without the SO term ( $V^{(\mathrm{h})}_{\mathrm{so}} = 0$) and (b) with SO included.
The first case corresponds to the pure Fock-Darwin model, the results are not only inaccurate from quantitative point of view, but also predict a wrong spin ordering. In particular, this leads to lack of the  pronounced anticrossing at $B_{z} \approx 10$~T. On the other hand, the full effective model (Fig.~\ref{fig:mag_p_h}b) gives correct spin ordering and very good agreement to the numerical $14$~$\kp$ data. The results of the fitting procedure are given in Tab.~\ref{tab:param_fit_h}. In contrast to the electron case, both anisotropy ($V^{(\mathrm{h})}_{\mathrm{a}}$) and spin-orbit parameter ($V^{(\mathrm{h})}_{\mathrm{so}}$) are larger in the case of the uniform QD. Identification of the most pronounced spin-orbit coupling mechanism for a given system gives a possibility to control this effect by a proper growth of the sample. For example, the abrupt material interfaces could be softened by annealing, shear strain could be reduced by using a strain reducing layer, and the strength of the Dresselhaus coupling could be (partially) controlled by the material composition. In order to assess various contributions to the overall spin-orbit coupling in the $p$-shell, I performed the calculation within several degrees of approximation:

\begin{enumerate}
 \item	This contains $V^{(\mathrm{h})}_{\mathrm{so}}$ obtained from the $14$ band $\kp$ model. All of the terms in the Hamiltonian are present. 
 \item  The full $8$ $\kp$. The Dresselhaus SO coupling (which arises from the coupling to $\Gamma_{8c}$ and $\Gamma_{7c}$) is accounted for via perturbative terms (see $H^{\mathrm{D}}$ in the Appendix). The model inherently contains also the Rashba coupling  (except some relatively small contributions from the coupling to $\Gamma_{8c}$ and $\Gamma_{7c}$) and other spin-orbit mechanisms related to material inhomogeneity at the interfaces. 
 \item  The same as (2), but with $d_{\mathrm{v}} = 0$ which neglects the influence of the shear strain in the valence band. 
 \item  The same as (3), with further reduction of spin-orbit coupling by disabling the Dresselhaus terms ($H^{\mathrm{D}} = 0$) and neglecting small contribution from $C_{\mathrm{k}}$ ($k$-linear terms). 
 \item  This approximation is based on $6$ $\kp$ model (valence bands only). The deformation potential $d_{\mathrm{v}}$ is present. The Dresselhaus coupling is neglected and  $C_{\mathrm{k}} = 0$. The model partially accounts for the Rashba coupling via position dependence of the $\kappa$ and $q$ parameters in the magnetic Hamiltonian ($H^{\mathrm{m}}$ in the Appendix). However, this contribution is overestimated due to lack of some terms with the opposite sign. 
 \item  The same as (5), but without the magnetic part of the Hamiltonian ($H^{\mathrm{m}}$), influence of the shear strain is neglected $d_{\mathrm{v}} = 0 $. The contribution from the interface-induced mixing of the heavy- and light- hole subbands\cite{durnev14} is removed by setting $\gamma_{3} = 0$.
\end{enumerate}
As shown in Tab.~\ref{tab:param_fit_hm}, for both QDs the most important contribution to $V^{(\mathrm{h})}_{\mathrm{so}}$ comes from the shear strain (which enters to the valence band block of the Hamiltonian with the deformation potential $d_{\mathrm{v}}$). The influence of the Dresselhaus terms is small. Furthermore, the results from $8$ and $14$ $\kp$ are in very good agreement, on the other hand $6$ $\kp$ overestimates the $p$-shell spin-orbit coupling by about $15$~\%.

\begin{table}[tb]
	\begin{tabular}{|c|l|c|c|}
		\hline 	
		\multirow{2}{*}{No.} & \multirow{2}{*}{Model} & \multicolumn{2}{c|}{Value of $V^{(\mathrm{h})}_{\mathrm{so}}$ (meV) }  \\ \cline{3-4}
		 &  & Uniform QD & Blurred QD \\
		\hline
		1. & 14 $\kp$ full  & $9.216$  & $7.047$ \\				
\hline 	
		2. & 8 $\kp$ full  & $9.276$ & $7.107$\\				
\hline 	 
		3. & 8 $\kp$, neglected $d_{\mathrm{v}} = 0$  & $3.195$ & $1.5381$\\				
\hline 	
\multirow{2}{*}{4.} & 8 $\kp$, neglected $d_{\mathrm{v}} = 0$, $C_{\mathrm{k}} = 0$  & \multirow{2}{*}{$3.180$} & \multirow{2}{*}{$1.5132$} \\
& and Dresselhaus terms $H^{\mathrm{D}} = 0$ &  &  \\			 	
\hline 		
		5. & 6 $\kp$ full  & $10.512$ & $8.1204$ \\
\hline 			
		 \multirow{2}{*}{6.} & 6 $\kp$, neglected $H^{\mathrm{m}} = 0$  & \multirow{2}{*}{0.0} & \multirow{2}{*}{0.0} \\
    & and $\gamma_{3} = 0$, $d_{\mathrm{v}} = 0$  &  &  \\				
		\hline 	
	\end{tabular} 
	\caption{\label{tab:param_fit_hm}The effective spin-orbit parameter $V^{(\mathrm{h})}_{\mathrm{so}}$ for hole $p$-shell obtained from various approximations. All of the parameter definitions are given in the Appendix.}
\end{table}

\begin{figure}[tb]
	\begin{center}
		\includegraphics[width=3.5in]{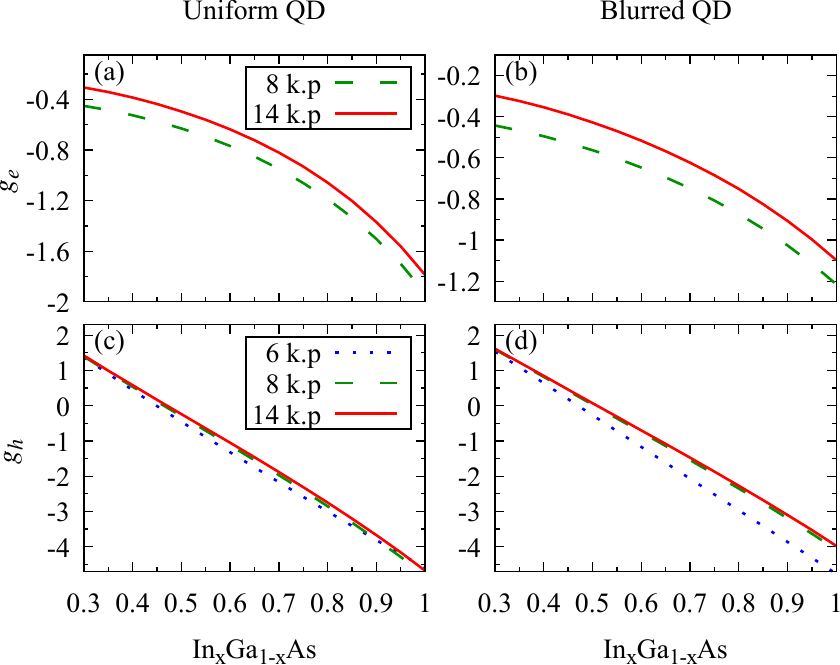}
	\end{center}
	\caption{\label{fig:compar}\textcolor{gray}({Color online) The electron (a,b) and hole (c,d) g-factor as a function of material composition in the case of uniform and blurred QD. } }
\end{figure}
I calculated electron and hole g-factors and compared the accuracy of various multiband approaches.  Fig.~\ref{fig:compar}(a,b) show electron g-factor as a function of the dot composition. In the absolute values, there is a good agreement between $8$ and $14$ band $\kp$, the relative discrepancy between them increases with decreasing InAs content (as the $g$-factor decreases). The comparison $8$ $\kp$ to $2$ $\kp$ is widely discussed in Ref.~\onlinecite{pyszczorski17a}.
Fig.~\ref{fig:compar}(c,d) show the hole $g$-factor obtained from $6$, $8$ and $14$~$\kp$ models. While, in the case of the uniform QD there is a good agreement between all of the methods, for the blurred QD, $6$ $\kp$ gives worse results. Due to the ellipticity condition in $14$ $\kp$, I reduced the optical matrix parameter $Q \rightarrow Q_{\mathrm{r}}$ (see the Appendix), where I took $80\%$ of the original value. As $Q$ contributes to the $g$-factors in the second power, the discrepancy between $8$ and $14$ could be larger by up to $36\%$ compared to the present values. Both for the electron and hole, the values for the blurred QD are significantly reduced, which can be related to enhanced GaAs content in the dot.

\section{Conclusions}
\label{sec:conclusion}
 I calculated the magnetic field dependence of the electron and hole states in a QD. I investigated the influence of spin-orbit coupling on the structure of electron and hole states. The results demonstrate an important impact of the shear strain on the spin-orbit coupling inside the hole $p$-shell.  I showed that numerical results can be very well accounted for by an empirical Fock-Darwin model if a term representing spin-orbit coupling is included. Finally, I compared the values of electron and hole g-factors obtained from $8$ and $14$ band $\kp$ calculation and showed that these methods are in reasonably good agreement. 

\acknowledgments
This work was supported by the
Polish National Science Centre (Grant No.~2014/13/B/ST3/04603). 
Calculations have been partly carried out in Wroclaw Centre for
Networking and Supercomputing (http://www.wcss.wroc.pl), Grant No. 203. 
I am grateful to Pawe{\l} Machnikowski for his helpful suggestions and to Micha{\l} Gawe{\l}czyk for sharing his implementation of the blur algorithm.

\appendix
\section{Calculation details} 

The kinetic part of the Hamiltonian can be expressed in terms of the crystal invariants\cite{winkler03}
\begin{align*}
H^{k}_{\mathrm{8c8c}} &= E'_{g} + \Delta'_{0},\\
H^{k}_{\mathrm{7c7c}} &= E'_{g},\\
H^{k}_{\mathrm{6c6c}} &= E_{g} + k_{x} \frac{\hbar^{2}}{2m'} k_{x} + \cp,\\
H^{k}_{\mathrm{8v8v}} &= -\frac{\hbar^{2}}{2m_{0}} \left \{ k_{x} \gamma'_{1}  k_{x} - 2   \left (  J^{2}_{x}  - \frac{1}{3} J^{2}  \right ) k_{x} \gamma'_{2}  k_{x}    \right . \\ 
&\phantom{=} - 4 \gamma'_{3}  \{J_{x},J_{y}\}  ( k_{x} \gamma'_{3} k_{y}  + k_{y} \gamma'_{3} k_{x})  + \cp \bigg \}\\
&\phantom{=} + \frac{1}{\sqrt{3}}  \left[ \{ J_{x},J^{2}_{y} - J^{2}_{z}  \} ( C_{k} k_{x} + k_{x} C_{k} )+ \cp  \right],\\
H^{k}_{\mathrm{7v7v}} &= -\Delta_{0} - \frac{\hbar^{2}}{2m_{0}}  k_{x} \gamma'_{1}  k_{x},\\	
H^{k}_{\mathrm{8c7c}} &= 0,\\
H^{k}_{\mathrm{8c6c}} &= -\sqrt{3} P' (U_{x} k_{x} + \cp ),\\
H^{k}_{\mathrm{8c8v}} &= - \frac{2}{3} Q_{\mathrm{r}} (\{ J_{y},J_{z} \} k_{x} + \cp ) + \frac{1}{3} \Delta^{-},\\
H^{k}_{\mathrm{8c7v}} &= -2 Q_{\mathrm{r}} (U_{yz} k_{x} + \cp ),\\
H^{k}_{\mathrm{7c6c}} &=  \frac{1}{\sqrt{3}} P' (\sigma_{x} k_{x} + \cp ),\\
H^{k}_{\mathrm{7c8v}} &=  -2 Q_{\mathrm{r}} (T_{yz} k_{x} + \cp ),\\
H^{k}_{\mathrm{7c7v}} &= -\frac{2}{3} \Delta^{-},\\
H^{k}_{\mathrm{6c8v}} &= \sqrt{3}  (T_{x} k_{x}   + \cp )P,\\	
H^{k}_{\mathrm{6c7v}} &= - \frac{1}{\sqrt{3}}  (\sigma_{x} k_{x}  + \cp ) P ,\\	
H^{k}_{\mathrm{8v7v}} &= -\frac{\hbar^{2}}{2m_{0}} \left[ -6  U_{xx} k_{x} \gamma'_{2} k_{x}  \right . \\ & \phantom{=}   -12  U_{xy} ( k_{x} \gamma'_{3} k_{y} + k_{y} \gamma'_{3} k_{x} ) + \cp \big ]	
\\ & \phantom{=} - i \frac{\sqrt{3}}{2}  \left [ U_{yz} ( C_{k} k_{x} + k_{x} C_{k} ) + \cp \right ],	
\end{align*}
where $E_{g},E'_{g}$ are the energy gaps between $\Gamma_{8v}$ -- $\Gamma_{6c}$ and $\Gamma_{8v}$ -- $\Gamma_{7c}$ respectively, $\Delta_{0},\Delta'_{0},\Delta^{-}$ are parameters related to the spin-orbit coupling, $P,P',Q_{\mathrm{r}}$ are proportional to the interband momentum matrix elements, $m_{0}$ is the free electron mass, $m'$ and $\gamma'_{i}$ are modified values of the effective mass and Luttinger parameters. The relations between the modified and the original parameters (which are listed in Tab.~\ref{tab:param}) are given further. The matrices $\sigma_{i}$ are Pauli matrices, $J_{i}$  are matrices of angular momentum,
$U_{i}$, $U_{ij}$ are hermitian conjugates of $T_{i}$, $T_{ij}$ respectively (which definitions are given in given in Refs.~\onlinecite{mayer91,winkler03,eissfeller12}). 
$k_{i}$ is represented as spacial derivatives in the real space. Then, a discretization is performed using the finite difference method\cite{andrzejewski10}. I avoid spurious solution problem related to the first order derivatives\cite{andlauerPhD} by applying the central $4$-point scheme. 
\begin{table}
	\begin{tabular}{llll}
		\hline
		& GaAs & InAs & bowing\\
		\hline
		\multirow{1}{*}{$m^{*}_{\mathrm{e}}$} & \multirow{1}{*}{$0.0665m_{0}$} & \multirow{1}{*}{$0.0229m_{0}$} & $ 0.0091 m_{0}$ \\	
		$E_{\mathrm{v}}$ & $0.0$~eV & $0.21$~eV & -\\
		$E_{\mathrm{g}}$ & 1.519~eV & 0.417~eV & $0.477$~eV\\
		$E'_{\mathrm{g}}$ & 4.488~eV & 4.390~eV & -\\
		\rule{0pt}{3ex}$P$ & \multicolumn{3}{c}{calculated  from $P = \sqrt{ E_{P}  \hbar^{2} / (2 m_{0} )}$}\\		
		$P'$ & $4.78 i$~$\mathrm{eV \AA}$ & $0.873 i$~$\mathrm{eV \AA}$ & -\\
		$Q$ & $8.165$~$\mathrm{eV \AA}$ & $8.331$~$\mathrm{eV \AA}$ & -\\		
		$Q_{\mathrm{r}}$ & \multicolumn{3}{c}{reduced value $Q_{\mathrm{r}} = 0.8 Q $}\\	
		$\gamma_{\mathrm{1}}$ & 6.98 & 20.0 & - \\
		$\gamma_{\mathrm{2}}$ & 2.06 & 8.5 & - \\
		$\gamma_{\mathrm{3}}$ & 2.93 & 9.2 & -\\  		
		$C_{\mathrm{k}}$ & -0.0034~$\mathrm{eV \AA}$ & -0.0112~$\mathrm{eV \AA}$ & -\\		
		$\Delta_{0}$ & 0.341~eV & 0.39~eV& $0.15$~eV \\ 
		$\Delta'_{0}$ & 0.171~eV & 0.24~eV& - \\ 		
		$\Delta'$ & -0.05~eV & 0.0& - \\ 
		$a_{\mathrm{c}}$ & -7.17~eV & -5.08~eV & $2.61$~eV\\
		$a_{\mathrm{v}}$ & 1.16~eV & 1.00~eV & -\\
		$b_{\mathrm{v}}$ & -2.0~eV & -1.8~eV & -\\
		$d_{\mathrm{v}}$ & -4.8~eV & -3.6~eV & -\\  
		$g$ & -0.44 & -14.9 & -\\  
		$\kappa$ & 1.2 & 7.6 & -\\ 
		$q$ & \multicolumn{3}{c}{calculated from \(\displaystyle q =  \frac{2 m_{0}}{\hbar^{2}} \frac{2}{9} \left (   \frac{Q^{2}}{E'_{\mathrm{g}}} - \frac{Q^{2}}{E'_{\mathrm{g}}+\Delta'_{0}}  \right )  \)}\\
	\end{tabular} 
	\caption{\label{tab:param}Material parameters used in the calculations.\cite{vurgraftman01,winkler03}}
\end{table} 

Strain enters into the Hamiltonian via terms\cite{winkler03}
\begin{align*}
H^{\mathrm{str}}_{\mathrm{6c6c}} &=  a_{\mathrm{c}} \Tr{\epsilon},\\
H^{\mathrm{str}}_{\mathrm{8v8v}} &=  a_{\mathrm{v}} \Tr{\epsilon} - b_{\mathrm{v}} \left[ \left( J^{2}_{x}  - \frac{1}{3} J^{2} \right)\epsilon_{xx} + \cp  \right] \\ 
& \phantom{=} - \frac{d_{\mathrm{v}}}{\sqrt{3}} \left[ 2 \{J_{x},J_{y}\} \epsilon_{xy}  + \cp \right],\\
H^{\mathrm{str}}_{\mathrm{7v7v}} &=  a_{\mathrm{v}} \Tr{\epsilon},\\
H^{\mathrm{str}}_{\mathrm{6c8v}} &=  - \sqrt{3}  \left( T_{x} \sum_{j=x,y,z} \epsilon_{xj} k_{j} + \cp \right) P, \\
H^{\mathrm{str}}_{\mathrm{6c7v}} &=  \frac{1}{\sqrt{3}} \left( \sigma_{x} \sum_{j=x,y,z} \epsilon_{xj} k_{j} + \cp \right) P, \\	
H^{\mathrm{str}}_{\mathrm{8v7v}} &=  -3 b_{\mathrm{v}} \left( U_{xx} \epsilon_{xx} + \cp \right) -  \sqrt{3} d_{\mathrm{v}} \left( 2 U_{xy} \epsilon_{xy} + \cp \right) ,
\end{align*}
where $a_{\mathrm{c}}$, $a_{\mathrm{v}}$, $b_{\mathrm{v}}$, $d_{\mathrm{v}}$ are deformation potentials. 
Magnetic interaction is accounted for via the elements\cite{eissfeller12} 
\begin{align*}
H^{\mathrm{m}}_{\mathrm{8c8c}} &=  \frac{g_{0}}{3} \mu_{\mathrm{B}} \bm J \cdot \bm B,\\
H^{\mathrm{m}}_{\mathrm{8c7c}} &=  - g_{0} \mu_{\mathrm{B}} \bm U \cdot \bm B,\\
H^{\mathrm{m}}_{\mathrm{7c7c}} &=  - \frac{g_{0}}{6} \mu_{\mathrm{B}} \bm \sigma \cdot \bm B,\\
H^{\mathrm{m}}_{\mathrm{6c6c}} &=  i \frac{\hbar^{2}}{4m_{0}} \left[  \left( k_{x} g' k_{y} - k_{y} g' k_{x}   \right)\sigma_{z} + \cp \right ],\\
H^{\mathrm{m}}_{\mathrm{8v8v}} &= -i \frac{\hbar^{2}}{m_{0}} \left[  \left( k_{x} \kappa' k_{y} - k_{y} \kappa' k_{x}   \right) J_{z} + \cp \right ] \\
& \phantom{=}  -i \frac{\hbar^{2}}{m_{0}} \left[  \left( k_{x} q' k_{y} - k_{y} q' k_{x}   \right) J^{3}_{z} + \cp \right ] ,\\
H^{\mathrm{m}}_{\mathrm{7v7v}} &= -i \frac{\hbar^{2}}{m_{0}} \left[  \left( k_{x} \kappa' k_{y} - k_{y} \kappa' k_{x}   \right) \sigma_{z} + \cp \right ] \\
&\phantom{=} - \mu_{\mathrm{B}} \bm \sigma \cdot \bm B,\\
H^{\mathrm{m}}_{\mathrm{8v7v}} &= -i \frac{3 \hbar^{2}}{2 m_{0}} \left[  \left( k_{x} \kappa' k_{y} - k_{y} \kappa' k_{x}   \right) U_{z} + \cp \right ] \\
                               &\phantom{=} - 3 \mu_{\mathrm{B}} \bm U \cdot \bm B,
\end{align*}
where $g_{0}=2$, $g'$, $\kappa'$ and $q'$ are related to the electron and hole g-factors. In the case of a nanostructure, the above elements are nonzero even at $B=0$ and $\kappa'$ introduces Burt-Foreman operator ordering which represents boundary conditions at the interface\cite{mlinar05,foreman97}. In the presence of magnetic field $k_{n} k_{m} - k_{m} k_{n} = -i \varepsilon_{nmk} e B_{k} / \hbar$, where $\varepsilon_{nmk}$ denotes Levi-Civita symbol.
For the calculation in $14$ band $\kp$, the contributions from $\Gamma_{\mathrm{8c}}$, $\Gamma_{\mathrm{7c}}$ and $\Gamma_{\mathrm{6c}}$ need to be removed. The modified parameters are then given by\cite{winkler03}
 \begin{align*}
 \frac{m_{0}}{m'} &=  \frac{m_{0}}{m^{*}_{\mathrm{e}}}  - \frac{2}{3} \frac{E_{P'}}{E_{g}-E'_{g}-\Delta'_{0}} - \frac{1}{3} \frac{E_{P'}}{E_{g}-E'_{g}} \\
 & \phantom{=} - \frac{2}{3} \frac{E_{P}}{E_{g}} - \frac{1}{3} \frac{E_{P}}{E_{g}+\Delta_{0}}  ,\\
  g' &=  g + \frac{2}{3} \frac{E_{P'}}{E_{g}-E'_{g}-\Delta'_{0}} - \frac{2}{3} \frac{E_{P'}}{E_{g}-E'_{g}} \\
  & \phantom{=} + \frac{2}{3} \frac{E_{P}}{E_{g}} - \frac{2}{3} \frac{E_{P}}{E_{g}+\Delta_{0}}  ,\\ 
  \gamma'_{1} &=  \gamma_{1} - \frac{1}{3} \frac{E_{\mathrm{Qr}}}{E'_{g}+\Delta'_{0}} - \frac{1}{3} \frac{E_{\mathrm{Qr}}}{E'_{g}} - \frac{1}{3} \frac{E_{P}}{E_{g}},\\    
  \gamma'_{2} &=  \gamma_{2} + \frac{1}{6} \frac{E_{\mathrm{Qr}}}{E'_{g}} - \frac{1}{6} \frac{E_{P}}{E_{g}},\\ 
  \gamma'_{3} &=  \gamma_{3} - \frac{1}{6} \frac{E_{\mathrm{Qr}}}{E'_{g}} - \frac{1}{6} \frac{E_{P}}{E_{g}},\\     
  \kappa' &=  \kappa - \frac{7}{18} \frac{E_{\mathrm{Qr}}}{E'_{g}+\Delta'_{0}} + \frac{5}{9} \frac{E_{\mathrm{Qr}}}{E'_{g}} - \frac{1}{6} \frac{E_{P}}{E_{g}},\\
  q' &=  q  - \frac{2}{9} \frac{E_{\mathrm{Qr}}}{E'_{g}} + \frac{2}{9} \frac{E_{\mathrm{Qr}}}{E'_{g}+\Delta'_{0}},  
 \end{align*}
where $E_{P} = 2 m_{0} P^{2}/\hbar^{2}$ is the Kane energy, analogously $E_{P'}  = 2 m_{0} \vert P' \vert ^{2}/\hbar^{2} $ and $E_{\mathrm{Qr}}  = 2 m_{0} Q_{\mathrm{r}}^{2}/\hbar^{2}$. In order to avoid spurious solutions related to losing the ellipticity\cite{xian10}, I reduce the values of $E_{P}$ and $Q$. In the first case, I use the rescaling relation\cite{birner11} 
\begin{displaymath}
	E_{P} = \left (\frac{m_{0}}{m^{*}_{\mathrm{e}}} - 1 \right) \frac{E_{\mathrm{g}} (E_{\mathrm{g}} + \Delta_{0})}{E_{\mathrm{g}} + 2\Delta_{0}/3}.
\end{displaymath}
For the latter, I reduce the parameter $Q \rightarrow Q_{\mathrm{r}}$ to $80$\% of the original value. Due to the inconsistency of the reported values\cite{winkler03,lawaetz71},  following Ref.\onlinecite{eissfeller12} I take $q$ from the perturbative formula (see Table~\ref{tab:param}). Although $14$ $\kp$ inherently describes the Dresselhaus coupling, in the case of the reduced $Q$ there is a need of compensation via perturbative formulas. The relevant part of the Hamiltonian is given by\cite{winkler03}
\begin{align*}
H^{D}_{\mathrm{6c8v}} &= i\frac{\sqrt{3}}{2} \left [ T_{x} 
(k_{y} B_{\mathrm{8v}}^{+} k_{z} + k_{z} B_{\mathrm{8v}}^{+} k_{y}) +\cp \right ] \nonumber \\
& \quad +\frac{\sqrt{3}}{2}
\left [ (T_{xx}-T_{yy}) \left (\frac{2}{3} k_{z}
B_{\mathrm{8v}}^{-} k_{z} \right . \right . \nonumber\\ 
& \quad  \left . - \frac{1}{3} k_{x} B_{\mathrm{8v}}^{-} k_{x} -
\frac{1}{3} k_{y} B_{\mathrm{8v}}^{-} k_{y} \right )\nonumber \\ 
& \quad-T_{zz}(k_{x} B_{\mathrm{8v}}^{-} k_{x} - k_{y}
B_{\mathrm{8v}}^{-} k_{y}) \bigg ], 
\label{6c8v} \\
H^{D}_{\mathrm{6c7v}} & 
= -\frac{i}{2\sqrt{3}} \left [ \sigma_{x}  ( k_{y} B_{\mathrm{7v}} k_{z} + k_{z} B_{\mathrm{7v}} k_{y} ) +\cp \right ],
\end{align*}
where
\begin{align*}
B^{+}_{\mathrm{8v}} &=  \frac{1}{2i} P' (Q-Q_{\mathrm{r}}) \left ( \frac{1}{E_{\mathrm{g}} - E'_{\mathrm{g}} - \Delta'_{0}} - \frac{1}{E'_{\mathrm{g}} + \Delta'_{0}} \right . \\
 & \quad  +  \left .	\frac{1}{E_{\mathrm{g}} - E'_{\mathrm{g}}} - \frac{1}{E'_{\mathrm{g}}} 			
	\right ),\\ 	
B^{-}_{\mathrm{8v}} &=  \frac{1}{2i} P' (Q-Q_{\mathrm{r}}) \left ( -\frac{1}{E_{\mathrm{g}} - E'_{\mathrm{g}} - \Delta'_{0}} - \frac{1}{E'_{\mathrm{g}} + \Delta'_{0}} \right . \\
& \quad  +  \left .	\frac{1}{E_{\mathrm{g}} - E'_{\mathrm{g}}} - \frac{1}{E'_{\mathrm{g}}} 			
\right ),\\  
B_{\mathrm{7v}} &=  \frac{1}{i} P' (Q-Q_{\mathrm{r}}) \left ( \frac{1}{E_{\mathrm{g}} - E'_{\mathrm{g}} - \Delta'_{0}} - \frac{1}{E'_{\mathrm{g}} + \Delta_{0} + \Delta'_{0}} \right ).	
\end{align*}
For the $8$ $\kp$ calculations, I perform the substitution $P' = 0$, $Q_{\mathrm{r}} = 0$, and $\Delta^{-}=0$ which decouples $\Gamma_{\mathrm{8c}}+\Gamma_{\mathrm{7c}}$ from the $\Gamma_{\mathrm{6c}} + \Gamma_{\mathrm{8v}} + \Gamma_{\mathrm{7v}}$ block. Further reduction by $P = 0$ and $H^{D} = 0$ gives $6$ $\kp$ Hamiltonian. 

The effective model of the carrier $p$-states can be written in a matrix representation. 
For holes, the basis $ \left \{ \ket{+1 \Uparrow},\ket{-1 \Uparrow},\ket{+1 \Downarrow},\ket{-1 \Downarrow} \right \}$  gives the matrix
\begin{align*}
H^{(h)}_{p,\mathrm{eff}} &= V^{\mathrm{(h)}}_{\mathrm{a}}
\left (  
\begin{array}{cccc} 
0& 1 & 0 & 0 \\
1 &0 & 0 & 0 \\
0& 0 & 0 & 1 \\
0& 0 & 1 & 0 \\
\end{array}
\right )  +  V^{\mathrm{(h)}}_{\mathrm{so}} \left (  
\begin{array}{cccc}
\frac{1}{2}  & 0 & 0 & 0 \\
0& -\frac{1}{2} & 0 & 0 \\
0& 0 & -\frac{1}{2}  & 0  \\
0& 0 & 0 & \frac{1}{2}  \\
\end{array}
\right )  \\
& \phantom{=} +  W^{\mathrm{(h)}}  B_{z}  \left (  
\begin{array}{cccc}
1  & 0 & 0 & 0 \\
0& 1 & 0 & 0 \\
0& 0 & -1  & 0  \\
0& 0 & 0 & -1  \\
\end{array}
\right ) + \alpha B^{2}_{z} \left (  
\begin{array}{cccc}
1  & 0 & 0 & 0 \\
0& 1 & 0 & 0 \\
0& 0 & 1  & 0  \\
0& 0 & 0 & 1 \\
\end{array}
\right ) \\
& \phantom{=}  + \mu_{B} g^{\mathrm{(h)}}_{\mathrm{p}} B_{z} \left (  
\begin{array}{cccc}
\frac{1}{2}  & 0 & 0 & 0 \\
0& -\frac{1}{2} & 0 & 0 \\
0& 0 & \frac{1}{2}  & 0  \\
0& 0 & 0 & -\frac{1}{2}  \\
\end{array}
\right ).
\end{align*}
This matrix can be splitted on two $2\times2$ matrices and diagonalized analytically with the eigenvalues
\begin{align*}
E_{1}(B_{z}) &=   \frac{1}{2} \mu_{B} g^{\mathrm{(h)}}_{\mathrm{p}} B_{z} \\
& \quad + \sqrt{ \left (W^{\mathrm{(h)}} B_{z}+\frac{1}{2} V^{\mathrm{(h)}}_{\mathrm{so}} \right )^{2} + (V^{\mathrm{(h)}}_{\mathrm{a}})^{2} }  + \alpha B^{2}_{z} \\
E_{2}(B_{z}) &=   -\frac{1}{2} \mu_{B} g^{\mathrm{(h)}}_{\mathrm{p}} B_{z} \\
& \quad + \sqrt{ \left (W^{\mathrm{(h)}} B_{z}-\frac{1}{2} V^{\mathrm{(h)}}_{\mathrm{so}} \right )^{2} + (V^{\mathrm{(h)}}_{\mathrm{a}})^{2} }  + \alpha B^{2}_{z} \\
E_{3}(B_{z}) &=   \frac{1}{2} \mu_{B} g^{\mathrm{(h)}}_{\mathrm{p}} B_{z} \\
& \quad - \sqrt{ \left (W^{\mathrm{(h)}} B_{z}+\frac{1}{2} V^{\mathrm{(h)}}_{\mathrm{so}} \right )^{2} + (V^{\mathrm{(h)}}_{\mathrm{a}})^{2} }  + \alpha B^{2}_{z} \\
E_{4}(B_{z}) &=   -\frac{1}{2} \mu_{B} g^{\mathrm{(h)}}_{\mathrm{p}} B_{z} \\
& \quad - \sqrt{ \left (W^{\mathrm{(h)}} B_{z}-\frac{1}{2} V^{\mathrm{(h)}}_{\mathrm{so}} \right )^{2} + (V^{\mathrm{(h)}}_{\mathrm{a}})^{2} }  + \alpha B^{2}_{z}.
\end{align*}
The same formulas can be obtained for the electron.


\bibliographystyle{prsty}
\bibliography{abbr,quantum2}
\end{document}